\documentclass[prl,twocolumn,showpacs,floatfix,amsmath,nofootinbib,amssymb,floatfix]{revtex4}
\usepackage{graphicx,color,dcolumn,booktabs,bm}
\usepackage{longtable,lscape}
\usepackage{txfonts}
\usepackage{overpic}
\usepackage{amssymb}
\usepackage{indentfirst}
\usepackage{feynmf}   
\usepackage{slashed}  
\usepackage{cases}
\usepackage{color}
\usepackage{multirow}
\usepackage{epstopdf}
\usepackage{float}
\usepackage{graphicx,color,dcolumn,booktabs,bm}
\usepackage[colorlinks, citecolor=blue,anchorcolor=red,menucolor=red, linkcolor=red,filecolor=red,runcolor=red,urlcolor=blue,frenchlinks=red]{hyperref}

\graphicspath{{Figures/}} %

\begin{document}

\title{Criterion for the Existence of the $G(3900)$ Resonance}

\author{Yin Huang$^{1,2}$}\email{huangy2019@swjtu.edu.cn}
\author{Xurong Chen$^{2,3,4,5}$} \email{xchen@impcas.ac.cn}
\affiliation{$^{1}$School of Physical Science and Technology, Southwest Jiaotong University, Chengdu 610031,China}
\affiliation{$^{2}$Southern Center for Nuclear-Science Theory (SCNT), Institute of Modern Physics, Chinese Academy of Sciences, Huizhou 516000, China}
\affiliation{$^{3}$Institute of Modern Physics, Chinese Academy of Sciences, Lanzhou 730000, China }
\affiliation{$^{4}$School of Nuclear Science and Technology, University of Chinese Academy of Sciences, Beijing 100049, China}
\affiliation{$^{5}$State Key Laboratory of Heavy Ion Science and Technology, Institute of Modern Physics, Chinese Academy of Sciences, Lanzhou 730000, China}

\begin{abstract}
About 18 years ago, the BABAR and the Belle Collaborations reported evidence for a new $c\bar{c}$ structure near $3900~\mathrm{MeV}$ in $e^{+}e^{-} \to D\bar{D}$, triggering debate over whether it represents a genuine hadronic state or arises from interference effects and the opening of the $D^{*}\bar{D}$ threshold. The issue remained unsettled until 2024, when BESIII, with an enlarged $e^{+}e^{-} \to D\bar{D}$ sample, confirmed the presence of the $G(3900)$ structure, reviving discussion of its nature. Although it has been proposed as a $P$-wave $D\bar{D}^{*}$ molecular state, no consensus has emerged, and interference effects remain a plausible explanation.
In this Letter, we show that analyzing the $e^{+}e^{-}\to D\bar{D}$ cross section as a function of the transverse momentum of the final-state $D$ meson reveals a prominent Jacobian peak. This feature cannot be generated by interference effects alone and thus provides direct evidence for the existence of $G(3900)$. Moreover, the peak position is highly sensitive to the actual mass of $G(3900)$, offering a precise determination method. The criterion illustrated by $G(3900)$ can be extended to other hadronic systems, offering a powerful tool to identify genuine resonances.
\end{abstract}

\maketitle
\textit{Introduction}--With the advancement of experimental techniques, an increasing number of new hadronic states have been observed~\cite{ParticleDataGroup:2024cfk}. In 2024, the BESIII Collaboration 
reported the observation of a novel hadronic structure, $G(3900)$, in the analysis of the process $e^{+}e^{-} \to D\bar{D}$~\cite{BESIII:2024ths}, as further discussed in a recent comprehensive overview 
of open-charm measurements~\cite{Wang:2025dur}. The measured parameters of this resonance are \( M_{G(3900)} = 3872.5 \pm 14.2 \pm 3.0\ \mathrm{MeV} \) and \( \Gamma_{G(3900)} = 179.7 \pm 14.1 \pm 7.0\ \mathrm{MeV} \), placing it remarkably close in mass to the well-known \( X(3872) \) resonance,
first discovered by the Belle Collaboration in 2003~\cite{Belle:2003nnu, Brambilla:2019esw, Chen:2022asf, Meng:2022ozq}. This discovery has attracted considerable attention and sparked further theoretical
investigations.

Due to its mass (central experimental value) being very close to the \( D\bar{D}^*/\bar{D}D^* \) threshold, some studies interpret it as a newly observed \( D\bar{D}^*/\bar{D}D^* \) molecular state.
For instance, the analysis presented in Ref.~\cite{Lin:2024qcq} interprets the structure as a prospective \( P \)-wave \( \bar{D}D^*/D\bar{D}^* \) molecule, based on solutions to complex-scaled Schr\"{o}dinger
equations incorporating one-boson exchange potentials. Additional theoretical support comes from Ref.~\cite{Ye:2025ywy}, where the authors construct an effective \( P \)-wave contact interaction between the
\( (D, D^*) \) doublet and its antiparticle under the heavy quark symmetry framework. By solving the Lippmann--Schwinger equation, they demonstrate the emergence of a dynamically generated pole near the
observed \( G(3900) \) position, independent of whether two or three charmonium-like states are introduced into the coupled-channel formalism.

Interestingly, early enhancements in the same mass region were reported by both the BABAR~\cite{BaBar:2006qlj} and the Belle~\cite{Belle:2007qxm} collaborations. At the time, these features were generally
attributed not to new hadronic states but to threshold effects associated with the opening of the \( D\bar{D}^* + D^*\bar{D} \) channels, as well as the nodal structure of the \( \psi(3S) \) radial wave function~\cite{Eichten:1979ms}. Subsequent analyses of existing data~\cite{BaBar:2006qlj, Belle:2007qxm} by Refs.~\cite{Uglov:2016orr, Nakamura:2023obk, Du:2016qcr} demonstrated that the observed line
shape in \( e^+e^- \to D\bar{D} \) cannot be satisfactorily described without explicitly including the contribution from \( G(3900) \), lending support to its interpretation as a distinct hadronic structure.
Nevertheless, some studies contend that these data alone are insufficient to conclusively establish \( G(3900) \) as a genuine resonance~\cite{Zhang:2009gy}. Even with the incorporation of the latest
experimental results~\cite{BESIII:2024ths}, recent analyses~\cite{Salnikov:2024wah, Husken:2024hmi} conclude that identifying \( G(3900) \) as a bona fide resonance remains uncertain, suggesting instead
that the observed enhancement may arise from rescattering effects, coupled-channel interference, or other non-resonant phenomena.

These divergent interpretations raise two pivotal questions:
\begin{itemize}
    \item Is \( G(3900) \) a genuine hadronic state?
    \item If so, can it be classified as a \( P \)-wave \( \bar{D}D^{*}/D\bar{D}^{*} \) molecular resonance?
\end{itemize}
In this Letter, we propose to study the cross section of the process \( e^{+}e^{-} \to D^{+}D^{-} \) as a function of the transverse momentum distribution of the final-state \( D \) meson.
The presence of \( G(3900) \) would manifest as a pronounced Jacobian peak, distinctly different from that caused by interference effects, thereby providing direct evidence for its existence.
However, our results do not exclude the interpretation of $G(3900)$ as a $D\bar{D}^{*}$ molecular state.  In the following, we detail the theoretical framework.

\textit{Our Strategies}--The differential cross section \( d\sigma/d\cos\theta \) for the reaction \( e^{+} e^{-} \to D \bar{D} \) can be measured with high precision, where \(\theta\) is the
angle between the final-state \(D\) meson and the beam axis, defined along the \(z\)-direction. From this measurement, one can extract the cross section projected onto the transverse momentum
of the \(D\) meson~\cite{Smith:1983aa}:
\begin{align}
\sigma^{-1} \frac{d\sigma}{d\mu} = \mu (1 - \mu^{2})^{-1/2} \, \sigma^{-1} \frac{d\sigma}{d\cos\theta}, \label{eq1}
\end{align}
where \(\mu^{2} = P_{\perp}^{2}/P^{2} = \sin^{2}\theta\), with \(P\) and \(P_{\perp}\) denoting the three-momentum and its transverse component of the final-state \(D\) meson, respectively.
Here, division by the total cross section \(\sigma\) serves to normalize the distribution.

Equation~\eqref{eq1} gives rise to a characteristic Jacobian peak at \(\mu = 1\), corresponding to the maximal transverse momentum of the \(D\) meson~\cite{Matchev:2019bon},
\begin{align}
P_{\perp}^{\max} = \sqrt{\lambda(M^{2}, m_{D}^{2}, m_{\bar{D}}^{2})}/(2 M),
\end{align}
where \(\lambda(x,y,z) = x^{2} + y^{2} + z^{2} - 2xy - 2xz - 2yz\) is the K\"{a}ll\'{e}n function, and \(M\) denotes the invariant mass of the intermediate state, including the \(G(3900)\)
resonance under study. If \(G(3900)\) exists as a genuine resonance, the associated Jacobian peak is expected at \(P_{\perp}^{\max} = 520.995\,\mathrm{MeV}\) and \(504.018\,\mathrm{MeV}\)
for the final states \(D^{0} \bar{D}^{0}\) and \(D^{+} D^{-}\), respectively.  In contrast, if the observed enhancement arises solely from interference between intermediate states or other
nonresonant effects, no such Jacobian peak should appear in this range.

Most importantly, Eq.~\ref{eq1} shows that the Jacobian peak is independent of the amplitude, which is used to calculate the differential cross section. Therefore, we employ the conventional
effective Lagrangian approach for the calculation.  Using the currently measured cross section for \( e^{+} e^{-} \to D \bar{D} \), we compute the normalized differential
cross section \(\sigma^{-1} d\sigma/d\mu\) both including and excluding the \(G(3900)\) contribution to assess the feasibility of experimentally distinguishing these scenarios.

\begin{figure}[http]
\begin{center}
\includegraphics[bb=80 620 1050 705, clip, scale=0.58]{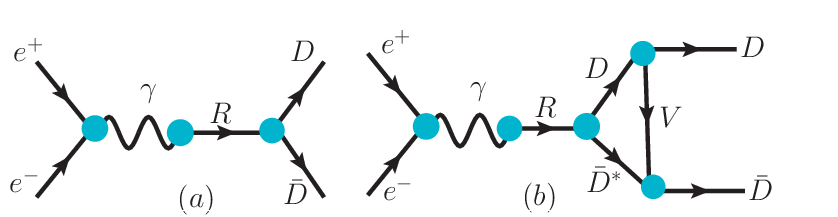}
\caption{The reaction \( e^{+} e^{-} \to D \bar{D} \) proceeds via intermediate resonances including \( R = \psi(3686), \psi(3770), \psi(4040),
\psi(4160) \), and \( G(3900) \) (a), as well as through the opening of the \( D \bar{D}^* \) channel (b). Here, \(V\) denotes the
exchanged vector meson.}
\label{cc1}
\end{center}
\end{figure}
We now evaluate $\mathrm{d}\sigma/\mathrm{d}\cos\theta$ in Eq.~\eqref{eq1}, treating $G(3900)$ as a genuine resonance and including the contributions from
$\psi(3686)$, $\psi(3770)$, $\psi(4040)$, $\psi(4160)$, as well as the opening of the $\bar{D}^*D$ threshold. Results without $G(3900)$ are obtained by simply
removing its contribution. The analysis focuses on the mass region $M_{G(3900)} = 3872.5 \pm 14.2 \pm 3.0$\,MeV, where these effects are most pronounced (Fig.~\ref{cc1}).
Note that, although $\psi(3686)$ lies about $200$\,MeV below $G(3900)$, its contribution is included because it can naturally account for the $e^{+}e^{-}\to D\bar{D}$
cross sections in the $3.7$--$3.8$\,GeV region~\cite{Li:2009pw}; otherwise, additional mechanisms would need to be invoked~\cite{Yang:2006nh}. Here we interpret the
$\psi(4040)$ contribution as a manifestation of the nodal structure of the $\psi(3S)$ radial wave function, as this effect is expected to correspond to a genuine $c\bar{c}$
state with a mass of about $4030$\,MeV~\cite{Eichten:1979ms}. The effects of the $\bar{D}^* D$ threshold opening, which contribute as resonances with $J^{PC} = 1^{--}$,
will be discussed in detail in the next section on Results and Discussion.

To evaluate the diagrams in Fig.~\ref{cc1}, we use the Lagrangians~\cite{Zhang:2009gy}
\begin{align}
\mathcal{L}_{V D \bar{D}} &= g_{V D \bar{D}} \left( D \partial_{\mu} \bar{D} - \partial_{\mu} D \bar{D} \right) V^{\mu}, \nonumber \\
\mathcal{L}_{V D \bar{D}^*} &= -i g_{V D \bar{D}^*} \varepsilon_{\alpha\beta\mu\nu} \partial^{\alpha} V^{\beta} \partial^{\mu} \bar{D}^{*\nu} D + \mathrm{H.c.},\label{eq3}
\end{align}
where \(P\) and \(V\) denote pseudoscalar and vector mesons, and \(\varepsilon_{\mu\nu\alpha\beta}\) is the antisymmetric tensor.
The couplings to vectors ($\rho$, $\omega$) in the chiral and heavy-quark limits~\cite{Cheng:2004ru} are
$g_{D^{*} D V} = \sqrt{2}\,\lambda\, g_{V}$ and $g_{D D V} = \beta g_{V}/\sqrt{2}$, with $g_{V} = m_{\rho}/f_{\pi}$, $f_{\pi}=132$\,MeV,
$\lambda=0.56$\,GeV$^{-1}$, and $\beta=0.59$~\cite{Zhang:2009gy,Liu:2006dq,Wise:1992hn}.  Here, $m_{\pi}$, $m_{\rho}$, $m_{D}$, and
$m_{D^{*}}$ are the meson masses.

For the $\psi/\gamma D D^{(*)}$ interactions we also employ the above Lagrangians.
The $\psi(3770) D\bar{D}$ coupling is fixed from the partial width,
$
\Gamma_{\psi(3770)\to D\bar{D}} = g_{\psi(3770) D\bar{D}}^{2} |\vec{p}\,|^3/(6\pi m_{\psi(3770)}^2),
$
which gives $g_{\psi(3770) D^{+} D^{-}}=13.41$ and $g_{\psi(3770) D^{0} \bar{D}^{0}}=12.65$~\cite{ParticleDataGroup:2024cfk}.
Other $\psi D^{(*)}\bar{D}^{(*)}$ couplings are fitted from data.
The $G(3900)$ production in $e^{+}e^{-}\to D\bar{D}$ is poorly constrained in the vector-meson dominance (VMD) framework~\cite{Bauer:1975bw}
and is also fitted from data.  In contrast, the $\psi(3686, 3770,4040,4160)$--$\gamma$ couplings are extracted via VMD from $\Gamma_{\psi\to e^+e^-}$
in the $m_e\to 0$ limit~\cite{Bauer:1975bw}, with $e/f_{\psi}=\left[3\,\Gamma_{\psi\to e^{+}e^{-}}/(2\,\alpha_{e}|\vec{p}_{e}|)\right]^{1/2}$,
where $|\vec{p}_e|$ is the electron momentum in the $\psi$ rest frame, and $\alpha_{e}=1/137$.

\begin{table*}[http!]
\centering
\caption{Fit parameters for $e^{+}e^{-}\to D\bar{D}^{(*)}$ with different models.
$\Gamma_{e^{+}e^{-}}$ denotes the $\psi(3900)\to e^{+}e^{-}$ width (keV).
``$\times$'' indicates a quantity not included in the corresponding model.
$g_{G(3900)/\psi D\bar{D}^{(*)}}$ represents the couplings of $\psi(3686)$, $\psi(3900)$, $\psi(4040)$, and $\psi(4160)$ to $D\bar{D}^{(*)}$ (GeV).
$\Lambda$ is the three-momentum cutoff (GeV).
$\theta_i$ $(i=1\!-\!5)$ are the amplitude phases of $\psi(3770)$, $\psi(3686)$, $\psi(4040)$, $\psi(4160)$, and $\psi(3900)/\psi(4009)$, respectively.
Models: Full includes $\psi(3900)$, $\psi(4040)$, and $\psi(4160)$;
$\psi(1)$ includes $\psi(4040)$ and $\psi(4160)$;
$\psi(2)$ includes only $\psi(4040)$;
$\psi(3)$ includes $\psi(4040)$ plus a background term.
The first four models are fitted to $e^{+}e^{-}\to D^{+}\bar{D}^{*-}$, while the remaining two are applied to $e^{+}e^{-}\to D^0\bar{D}^0$.}\label{tab-1}
\setlength{\tabcolsep}{-3.8mm}{
\begin{tabular}{ccccccccccccccccc}
\hline\hline
 ~~~~~~Model    &~~~~~~~~~~~ $\Gamma_{e^{+}e^{-}}$  &~~~~~~~~~~~~~$g_{G(3900)D\bar{D}^{(*)}}$&~~~~~~~~~~~ $g_{\psi(4040)D\bar{D}^{(*)}}$ &~~~~~~~~~~~~~ $g_{\psi(4160)D\bar{D}^{(*)}}$  &~~~~~~~~~~
                $g_{\psi(3686)D\bar{D}^{(*)}}$
              &~~~~~~~~~~~~$\Lambda$              &~~~~~~~~~~~~~$\theta_1$                    &~~~~~~~~~~~~$\theta_2$             &~~~~~~~~~~~~~~~$\theta_3$               &~~~~~~~~~~~~~$\theta_4$               &~~~~~~~~~~~~~$\theta_5$            &~~~~~~~~~~~~~~{\scriptsize $\chi^2/dof$}   ~~~~~~  \\ \hline
 ~~~~~~$Full$   &~~~~~~~~~ $152.437$               &~~~~~~~~~~~~~~~$0.334$                     &~~~~~~~~~~~ $0.863$                 &~~~~~~~~~~~~~ $0.152$                   &~~~~~~~~~~ $\times$
              &~~~~~~~~~~~~$\times$               &~~~~~~~~~~~~~$\times$                    &~~~~~~~~~~~~ $\times$              &~~~~~~~~~~~~~~~$0$                        &~~~~~~~~~~~~~$2.223^{\circ}$                  &~~~~~~~~~~~~~$43.031^{\circ}$      &~~~~~~~~~~~~~ 1.628                      ~~~~~~  \\
 ~~~~~~$\psi(1)$ &~~~~~~~~~ $\times$              &~~~~~~~~~~~~~~~$\times$                  &~~~~~~~~~~~ $1.546$                  &~~~~~~~~~~~~~ $0.395$                    &~~~~~~~~~~ $\times$
              &~~~~~~~~~~~~$\times$               &~~~~~~~~~~~~~$\times$                    &~~~~~~~~~~~~ $\times$              &~~~~~~~~~~~~~~~$0$                       &~~~~~~~~~~~~~$151.341^{\circ}$                  &~~~~~~~~~~~~~$\times$              &~~~~~~~~~~~~~ 4.794                      ~~~~~~  \\
 ~~~~~~$\psi(2)$ &~~~~~~~~~ $\times$              &~~~~~~~~~~~~~~~$\times$                  &~~~~~~~~~~~ $1.539$                  &~~~~~~~~~~~~~ $\times$                   &~~~~~~~~~~ $\times$
              &~~~~~~~~~~~~$\times$               &~~~~~~~~~~~~~$\times$                    &~~~~~~~~~~~~ $\times$              &~~~~~~~~~~~~~~~$0$                       &~~~~~~~~~~~~~$\times$                  &~~~~~~~~~~~~~$\times$              &~~~~~~~~~~~~~ 5.824                      ~~~~~~  \\
 ~~~~~~$\psi(3)$ &~~~~~~~~~ $\times$              &~~~~~~~~~~~~~~~$\times$                  &~~~~~~~~~~~ $1.087$                  &~~~~~~~~~~~~~ $\times$                   &~~~~~~~~~~ $\times$
              &~~~~~~~~~~~~$\times$               &~~~~~~~~~~~~~$\times$                    &~~~~~~~~~~~~ $\times$              &~~~~~~~~~~~~~~~$0$                       &~~~~~~~~~~~~~$\times$                  &~~~~~~~~~~~~~$\times$              &~~~~~~~~~~~~~ 2.742                      ~~~~~~  \\
 ~~~~~~$\psi(4)$ &~~~~~~~~~ $\times$              &~~~~~~~~~~~~~~~$\times$                  &~~~~~~~~~~~ $\times$                  &~~~~~~~~~~~~~ $\times$                 &~~~~~~~~~~ $\times$
              &~~~~~~~~~~~~$\times$               &~~~~~~~~~~~~~$\times$                    &~~~~~~~~~~~~ $\times$              &~~~~~~~~~~~~~~~$\times$                   &~~~~~~~~~~~~~$\times$                  &~~~~~~~~~~~~~$\times$              &~~~~~~~~~~~~~ 1.23                       ~~~~~~  \\\hline
 ~~~~~~$A_1$  &~~~~~~~~~ $152.437$                &~~~~~~~~~~~~~~~  $0.129$                 &~~~~~~~~~~~ $0.415$                  &~~~~~~~~~~~~~ $0.219$                   &~~~~~~~~~~ -5.351
              &~~~~~~~~~~~~$1.372\times{10}^{-5}$ &~~~~~~~~~~~~~$27.789^{\circ}$            &~~~~~~~~~~~~~~ $0^{\circ}$           &~~~~~~~~~~~~~~~$34.384^{\circ}$          &~~~~~~~~~~~~~$87.986^{\circ}$                   &~~~~~~~~~~~~~$124.273^{\circ}$     &~~~~~~~~~~~~~ 1.120                      ~~~~~~  \\
  ~~~~~~$A_{2}$&~~~~~~~~~ $9.312$             &~~~~~~~~~~~~~~~  $1.655$                     &~~~~~~~~~~~ $0.415$                  &~~~~~~~~~~~~~ $0.220$                   &~~~~~~~~~~ -5.351
              &~~~~~~~~~~~~$1.417\times{10}^{-6}$ &~~~~~~~~~~~~~$27.805^{\circ}$            &~~~~~~~~~~~~~~ $0^{\circ}$           &~~~~~~~~~~~~~~~$34.389^{\circ}$          &~~~~~~~~~~~~~$88.016^{\circ}$                   &~~~~~~~~~~~~~$124.224^{\circ}$     &~~~~~~~~~~~~~ 1.119                      ~~~~~~  \\
 ~~~~~~$A_3$  &~~~~~~~~~ $\times$                 &~~~~~~~~~~~~~~~  $\times$                &~~~~~~~~~~~ $1.218$                   &~~~~~~~~~~~~~ $0.116$                  &~~~~~~~~~~ -6.087
              &~~~~~~~~~~~~~~0.0357               &~~~~~~~~~~~~~$26.354^{\circ}$            &~~~~~~~~~~~~~~~$0^{\circ}$            &~~~~~~~~~~~~~~~$37.959^{\circ}$        &~~~~~~~~~~~~~$39.287^{\circ}$                    &~~~~~~~~~~~~~$22.725^{\circ}$      &~~~~~~~~~~~~~ 1.189                      ~~~~~~   \\
\hline \hline
\end{tabular}}
\end{table*}
Then, the amplitudes corresponding to Fig.~\ref{cc1} read
\begin{align}
{\cal M}_{a} &= \sum_{\psi_{j}}g_{\psi_{j} D \bar{D}}\,{\cal{K}}^{\nu}_{\psi_j} (p_{1} - p_{2})_{\nu},
\label{eq:Ta}\\
{\cal M}_{b} &= \sum_{\psi_{j}}i{\cal{K}}^{\nu}_{\psi_j}{\cal B}_{\psi_j}\sum_{V}\int \frac{d^{4}k}{(2\pi)^{4}}\epsilon_{\alpha\nu\sigma\eta}q^{\alpha}l_1^{\sigma}\epsilon_{\tau\varpi\theta\lambda}k^{\tau}l_1^{\theta}\nonumber\\
&\times{}\frac{g^{\eta\lambda}-l_1^{\eta}l_1^{\lambda}/m^2_{\bar{D}^{*}}}{l_1^{2} - m_{D^{*}}^{2}}\frac{g^{\varpi\xi}-k^{\varpi}k^{\xi}/m_V^2}{k^{2} - m_{V}^{2}}
\frac{(2p_1^{\xi}+k^{\xi})}{(p_{1}+k)^{2} - m_{D}^{2}}\label{eq:Ta-ty},
\end{align}
with
\begin{align}
{\cal{K}}^{\nu}_{\psi_j}=\frac{e^{2}}{q^{2}}\,\bar{v}(k_{2})\gamma_{\mu}u(k_{1})
\frac{m_{\psi_{j}}^{2}}{f_{\psi_{j}}}
\frac{g^{\mu\nu} - q^{\mu}q^{\nu}/m_{\psi_{j}}^{2}}
     {q^{2} - m_{\psi_{j}}^{2} + i\,m_{\psi_{j}}\Gamma_{\psi_{j}}}e^{i\theta_j},
\end{align}
Here $\theta_j$ denotes the phase of the amplitude associated with the corresponding resonance $\psi_j$.
${\cal B}_{\psi_i} = g_{V DD^{*}}\,g_{\psi_i D \bar{D}^{*}}\,g_{V D \bar{D}}$, $l_1 = p_2 - k$, and
$q^2 = (k_1 + k_2)^2 = (p_1 + p_2)^2 \equiv w^2$ is the squared total c.m.energy. $\Gamma_{\psi_j}$ denotes
the total width of the charmonium resonance, $u(k_1)$ and $v(k_2)$ are the electron and positron spinors,
and $p_1$, $p_2$ are the final-state $D$ and $\bar{D}$ momenta. To regularize the loop integrals, we employ a three-momentum cutoff.
Specifically, the integral in Eq.~\ref{eq:Ta-ty} is expanded as
$
\int d^4 k = \int dk^0\, |\vec{k}|^2\, d|\vec{k}|\, d\cos\theta\, d\phi ,
$
with $k^0 \in (-\infty,\infty)$, $|\vec{k}| \in (0, \Lambda)$, $\cos\theta \in (-1,1)$, and $\phi \in (0,2\pi)$.
The $k^0$ integral is evaluated using the residue theorem. The three-momentum cutoff $\Lambda$
applies to the $|\vec{k}|$ integration and is determined by fitting to experimental data.

Thus, the total transition amplitude is ${\cal M}_{fi} = {\cal M}_a + {\cal M}_b$, and the differential
cross section in the c.m. frame reads
\begin{equation}
\frac{d\sigma}{d\cos\theta} = \frac{1}{32\pi s} \frac{|\mathbf{p}_f|}{|\mathbf{p}_e|} \left|{\cal M}_{fi}\right|^2,
\label{eq:diffcs}
\end{equation}
where $\mathbf{p}_e$ and $\mathbf{p}_f$ are the three-momenta of the initial electron (or positron) and the final $D$
meson, respectively. Neglecting the electron mass, $|\mathbf{p}_e| = w/2$.

\textit{Results and Discussions}-- We discuss the effects of the $D\bar{D}^*$ threshold opening based on existing $e^{+} e^{-} \to D\bar{D}^*$
data~\cite{Belle:2006hvs}.  Previous studies~\cite{Zhang:2009gy} have shown that describing this effect solely through direct virtual-photon coupling is
insufficient, and resonant contributions must also be included.  For instance, a resonance with $m = 3.943 \pm 0.014~\mathrm{GeV}$ and $\Gamma = 119 \pm 10~\mathrm{MeV}$
was identified in a fit~\cite{Zhang:2009gy}, which, however, does not correspond to the $G(3900)$ observed by BESIII~\cite{BESIII:2024ths}.  In this energy
region, both $\psi(4040)$ and $\psi(4160)$ are known to decay into $D\bar{D}^*$~\cite{ParticleDataGroup:2024cfk}, and their production followed by decay into
$D\bar{D}^*$ can serve as a source of the threshold-opening effect. As studies suggest, if the $G(3900)$ indeed exists, it can be interpreted as a $P$-wave $D\bar{D}^{*}$
molecular state~\cite{Lin:2024qcq,Ye:2025ywy}, and thus it is expected to contribute to the $e^{+}e^{-} \to D\bar{D}^{*}$ reaction.
Consequently, the $G(3900)$ should also be considered as a source of the $D\bar{D}^{*}$ threshold-opening effect.

\begin{figure*}[http]
\begin{center}
\includegraphics[bb=20 10 2000 450, clip, scale=0.28]{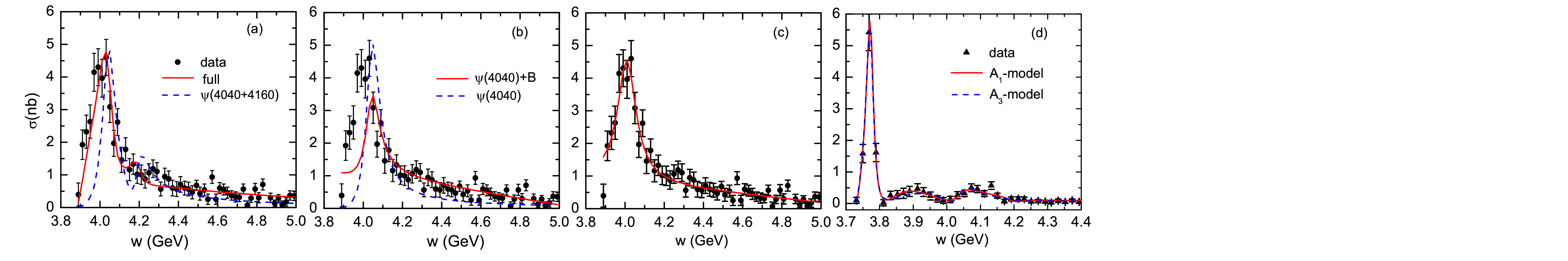}
\caption{Cross section of $e^{+} e^{-} \to D^{+} \bar{D}^{*-}/D^0\bar{D}^0$ (black dots/trangle) measured by the Belle~\cite{Belle:2006hvs} and the Belle~\cite{Belle:2007qxm}, respectively.
(a) Fit including $\psi(4040)$, $\psi(4160)$, and $G(3900)$ (red solid) versus only $\psi(4040)$ and $\psi(4160)$ (blue dashed).
(b) $\psi(4040)$ alone (blue dashed) and with a background $3.701 - 0.922 \,(w-1.052)$ (red solid).
(c) Fit with an additional unknown particle plus a background similar to (b) (red solid). (d) The red solid line and the blue dashed line correspond to the fit
results obtained by considering different resonant contributions, representing the $A_1$ and $A_3$ models,
respectively.}
\label{cc2}
\end{center}
\end{figure*}

However, the Belle~\cite{Belle:2006hvs} and the BABAR~\cite{BaBar:2009elc} measurements show that
$e^{+}e^{-} \to D\bar{D}^{*} + \mathrm{c.c.}$ is dominated by $\psi(4040)$, with no clear evidence for $X(3900)$ or $\psi(4160)$.
Surprisingly, our analysis shows that including the $G(3900)$, $\psi(4040)$, and $\psi(4160)$ with their experimental central
values provides an excellent description of the $e^{+}e^{-} \to D\bar{D}^{*}$ data (Fig.~\ref{cc2}(a)).
This suggests the presence of the $G(3900)$ in the $e^{+}e^{-} \to D\bar{D}^{*}$ process.
By contrast, considering only $\psi(4040)$ or $\psi(4040)$ and $\psi(4160)$  gives an even worse fit,
as quantified by the $\chi^2$ values in Table~\ref{tab-1}, where other fit parameters are also listed.

As shown in Table~\ref{tab-1}, the $\psi(2)$ model, which includes only the $\psi(4040)$ contribution, yields the
largest $\chi^2/\mathrm{dof} = 5.824$ and fails to describe the data (Fig.~\ref{cc2}(b)), in contrast to Belle~\cite{Belle:2006hvs}
and BABAR~\cite{BaBar:2009elc}, which suggest a $\psi(4040)$-dominant behavior.
The $\psi(1)$ model, incorporating both $\psi(4040)$ and $\psi(4160)$ contributions, likewise fails to provide a satisfactory
fit. Following Ref.~\cite{Zhang:2009gy}, the inclusion of a background term $3.701 - 0.922 \,(w-1.052)$, potentially arising
from direct $\gamma^* D\bar{D}^{*}$ coupling, significantly reduces $\chi^2/\mathrm{dof}$ and improves the fit (Fig.~\ref{cc2}(b)
and the $\psi(3)$ model in Table~\ref{tab-1}).
While adding this background to the $\psi(1)$ model enhances its performance, it still underperforms relative to the present
approach. The best fit, denoted as the $\psi(4)$ model, is achieved by introducing an additional particle with $m = 4009~\mathrm{MeV}$ and
$\Gamma = 142.43~\mathrm{MeV}$, together with a background term $0.6254 - 0.6445 (w - 4.3)$, yielding $\chi^2/\mathrm{dof} = 1.23$ (Fig.~\ref{cc2}(c)).
In this scenario, $\Gamma_{\psi(4009)\to e^{+}e^{-}} = 2.811~\mathrm{keV}$, and the couplings to $D\bar{D}^{*}$ exhibit a $w$-dependence,
consistent with Ref.~\cite{Zhang:2009gy}. Therefore, we use the full and $\psi(4)$ models as representative $D\bar{D}^{*}$ threshold effects
to assess their respective impacts.

We now present results treating $G(3900)$ as a genuine resonance in $e^{+}e^{-}\to D\bar{D}$. In this scenario, the $full$ and $\psi(4)$ models are considered separately
as representative contributions of the $D\bar{D}^{*}$ threshold effects. Focusing on the $full$ model, which includes the $G(3900)$ contribution in $e^{+}e^{-}\to D\bar{D}^{*}$,
the corresponding fit is shown as the red solid line in Fig.~\ref{cc2}(d), with parameters listed under the $A_1$ model in Table~\ref{tab-1}. For simplicity, both the tree-level
and loop contributions each include an amplitude with zero phase, corresponding to $\theta_3$ and $\theta_2$, respectively.  The overall contribution of the $full$ model is minimal.

In this fit, treating $G(3900)$ as a genuine resonance yields $\chi^2/\mathrm{dof} = 1.120$, providing an excellent description of the data. The $G(3900)$ width into $e^{+}e^{-}$
is $152.437~\mathrm{keV}$, whereas its decay into $D\bar{D}$-already observed experimentally-is only $38.513~\mathrm{keV}$, obtained by substituting $g_{G(3900)D\bar{D}}=0.129~\mathrm{GeV}$
into Eq.~\eqref{eq3}. Notably, interpreting $G(3900)$ as a $D\bar{D}^{*}$ molecule would imply that the $D\bar{D}$ width is 3.958 times smaller than that into $e^{+}e^{-}$, possible
only if $G(3900)$ contributes negligibly to $e^{+}e^{-}\to D\bar{D}^{*}$. A possible interpretation of this discrepancy is that a $D\bar{D}^{*}$ molecular assignment for $G(3900)$ may be inappropriate.

Including the $\psi(4)$ model, which accounts for the new particle and background in the $e^{+}e^{-}\to D\bar{D}^{*}$ fit, yields a slightly better description than the
$A_1$ model, with $\chi^2/\mathrm{dof} = 1.119$ versus 1.120. The linear shapes are nearly indistinguishable, so the $A_2$ results are omitted from Fig.~\ref{cc2}.
Compared with the $A_1$ model, the results are rather conventional: the $G(3900)$ widths into $D\bar{D}$ and $e^+e^-$ are 1.573~MeV and 9.312~keV, respectively, indicating
that $D\bar{D}$ is not the dominant decay channel. This, in turn, allows for the interpretation of $G(3900)$ as a $D\bar{D}^{*}$ molecular state, which contradicts the 
conclusion of the $A_1$ model that excludes a $D\bar{D}^{*}$ interpretation of $G(3900)$ based on the ratio of its decay widths into $D\bar{D}$ and $e^{+}e^{-}$. This 
again highlights the strong parameter dependence of fits to experimental data, where different reasonable parameter choices can lead to qualitatively different conclusions, 
preventing a unified interpretation at present. This situation also reflects the broader reason for the ongoing controversy, namely that most conclusions drawn from fits to 
experimental data exhibit significant model and parameter dependence.  Moreover, the $\psi(4)$ contribution is negligible due to $\Lambda = 1.417\times 10^{-6}$~GeV, with 
the main contributions coming from $\psi(3686)$, $\psi(3770)$, $G(3900)$, $\psi(4040)$, and $\psi(4160)$ via their tree-level couplings to $D\bar{D}$.

This effect becomes pronounced when $G(3900)$ is not treated as a genuine resonance, and the enhancement in the $e^{+}e^{-}\to D\bar{D}$ cross section in the energy region
without $G(3900)$ can be attributed to interference with other contributions~\cite{Salnikov:2024wah, Husken:2024hmi}. Our fits show that the $\psi(4)$ model, together with
tree-level contributions from $\psi(3686)$, $\psi(3770)$, $\psi(4040)$, and $\psi(4160)$, describes the $e^{+}e^{-}\to D\bar{D}$ data within experimental uncertainties, even
in the region where the $A_1$ and $A_2$ models require a $G(3900)$ contribution. The fit, shown as the blue dashed line in Fig.~\ref{cc2}(d), uses parameters listed under
the $A_3$ model in Table~\ref{tab-1}. This fit requires $\Lambda = 0.0357~\mathrm{GeV}$, yielding $\chi^{2}/\mathrm{dof} = 1.189$, where $\theta_{5}$ denotes the phase of
the $\psi(4009)$ production amplitude.

\begin{figure}[http]
\begin{center}
\includegraphics[bb=-120 8 1050 391, clip, scale=0.30]{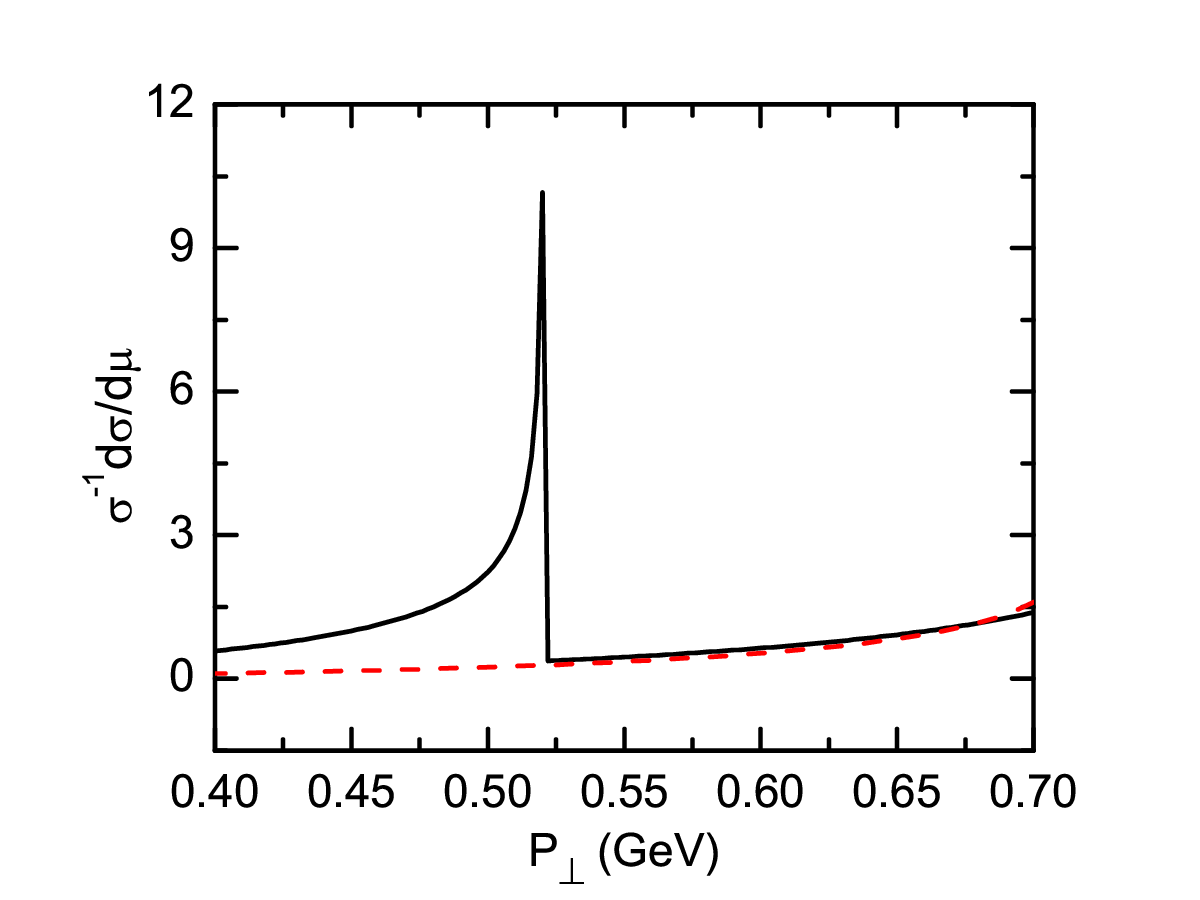}
\caption{The cross section of the $e^{+}e^{-}\to D^0\bar{D}^0$ process as a function of the transverse momentum $P_{\perp}$ of the final-state $D$ meson at $w=3872.5$ MeV.
The black solid line corresponds to the case with the $G(3900)$ contribution, while the red dashed line denotes the case without $G(3900)$.}
\label{cc3}
\end{center}
\end{figure}

All three models, $A_1$, $A_2$, and $A_3$, describe the $e^{+}e^{-}\to D\bar{D}$ data well, which are measured only via the two-body invariant mass spectrum of the final-state
$D\bar{D}$. This makes it difficult to establish $G(3900)$ as a genuine resonance. Crucially, by measuring the $e^{+}e^{-}\to D\bar{D}$ cross section as a function of the
transverse momentum of the final-state $D$ mesons, one can discriminate whether $G(3900)$ exists as a true resonance. If $G(3900)$ is genuine, it produces a Jacobian peak at
$P_{\perp}^{\rm max} = 520.995~\mathrm{MeV}$ and $504.018~\mathrm{MeV}$ for $D^{0}\bar{D}^{0}$ and $D^{+}D^{-}$, respectively. Using the amplitude fitted to the experimental
data, we find a clear Jacobian peak at $w = 3872.5~\mathrm{MeV}$ ($P_{\perp}^{\rm max} = 520.995~\mathrm{MeV}$), shown as the black solid line in Fig.~\ref{cc3}. In contrast,
when $G(3900)$ is not treated as a genuine resonance, this peak disappears, as indicated by the red dashed line.

The Jacobian peak originates from the $\mu (1-\mu^2)^{-1/2}$ term in Eq.~\ref{eq1}, reflecting a kinematic boundary~\cite{Matchev:2019bon}. Its position is uniquely fixed by
the mass of $G(3900)$, independent of model parameters. The distribution rises sharply as $P_{\perp}$ approaches its kinematic limit. However, in Fig.~\ref{cc3}, the spectrum
extends beyond the peak at $P_{\perp}=0.521~\mathrm{GeV}$ up to 0.7~GeV. This extension is caused by contributions from $\psi(4040)$ and $\psi(4160)$.  The strong sensitivity
of the peak position to the actual mass of $G(3900)$ makes it a promising tool for precision determination.

\textit{Summary}--- We study the properties of the recently observed $G(3900)$ in $e^+e^- \to D\bar{D}$ reactions. By analyzing the transverse momentum distribution of the 
final-state $D$ mesons, we identify a characteristic Jacobian peak, which provides a direct signature of the resonance nature of $G(3900)$. A comprehensive fit combining 
charmonium resonance contributions with $D\bar{D}^{(*)}$ threshold effects achieves an excellent description of the experimental data. This work establishes a robust strategy 
to distinguish genuine resonances from interference effects in hadron spectroscopy, thereby improving our understanding of exotic hadronic states.
It is worth noting that the model exhibits a strong dependence on the choice of parameters. As a consequence, it is not possible to unambiguously determine whether $G(3900)$ 
corresponds to a $P$-wave $D^{*}\bar{D}/\bar{D}^{*}D$ molecular state. Nevertheless, this ambiguity does not affect the appearance of the Jacobian peak, which remains a 
model-independent observable feature.

\section*{Acknowledgments}
This work is supported by National Key R\&{}D Program of China No.2024YFE0109800 and 2024YFE0109802.  Yin. Huang also acknowledges the support from the Fundamental Re
search Funds for the Central Universities under Grant No.
2682026TPY011.

\end{document}